\documentclass[11pt,twoside]{article}

\usepackage{asp2006}
\usepackage{epsf}
\usepackage{psfig}
\usepackage{lscape}

\begin{document}
\title{Data in the ADS -- Understanding How to Use it Better}
\author{Carolyn S. Grant, Alberto Accomazzi, Donna Thompson, Edwin Henneken, G\"unther Eichhorn, Michael J. Kurtz, and Stephen S. Murray}
\affil{Harvard-Smithsonian Center for Astrophysics, 60 Garden Street, Cambridge, MA 02138}

\begin{abstract}
The Smithsonian/NASA ADS Abstract Service contains a wealth of data for
astronomers and librarians alike, yet the vast majority of usage consists
of rudimentary searches.   Hints on how to obtain more focused search
results by using more of the various capabilities of the ADS are presented,
including searching by affiliation.  We also discuss the classification of 
articles by content and by referee status.

The ADS is funded by NASA Grant NNG06GG68G-16613687.
\end{abstract}

%%% MAIN BODY OF TEXT GOES HERE. CONSULT "INSTRUCTIONS FOR AUTHORS USING
%%% LATEX2E MARKUP", SECTIONS 2.3-2.6 FOR HELP WITH EQUATIONS, FIGURES,
%%% AND TABLES.

\section{Introduction}
Although the Smithsonian/NASA Astrophysics Data System (ADS) is used practically daily by most working astronomers
and used regularly by most working astronomy librarians, we have found
that there is not significant usage beyond basic searching by the large
majority of ADS users.  While it is certainly the case that basic
searching must satisfy a percentage of these users, it is also very likely
the case that some of them would benefit from understanding
how to make better use of the ADS in their searching.

We have done several things in an effort to educate people about how to
improve their use of the ADS;  we hand out ``hints on better use of the
ADS," we give talks, and we present posters.  Additionally,
we introduced the myADS notification service with the intention of customizing searches to user's
specific interests.  We have also made software changes such as the
improved author searching implemented in the beginning of 2006
to try to anticipate better what the user really wants.
While none of these things alone seems to be changing general ADS usage, we 
believe that gradually our user base is becoming better educated on 
modifying their use of the ADS.  

\section{Data Holdings in the ADS}

Since hitting one million
abstracts in January 1998, the number of abstracts in the ADS has grown 
steadily over time.  Likewise, since the initial purchase of citations from 
the Institute for Scientific Information (ISI) in 1999, the citations have 
also continuously climbed, particularly since 2002 when many of the journals 
began providing us with references in electronic format.

As of September 2006, the ADS contained a total of 4.87 million abstracts divided
into four databases: Astronomy (1.2 million), Physics (3.04 million), 
Preprints (0.38 million), and General Science (0.38 million).   Of these, 3.20
million contain abstracts (66\%) and 1.61 million have references (34\%).
In addition, the ADS contained 20.3 million citation pairs, 3.3 million
scanned pages, and 4.9 million external links as of September 2006.  

\section{More Effective Searching in the ADS}

Analysis of our data logs shows that the large majority of users perform
simple author queries to find their papers of interest.  Therefore,
for a typical user to gain more effective searching from the ADS, we need
to teach them how to improve their current search mode.  We do this
by broadcasting typical search hints and by advertising the usefulness of the 
myADS alerting service to inform users of new articles by those 
authors or about those topics in which they are most interested.

The myADS Update Service is our free custom
notification service promoting current awareness of the recent
technical literature in astronomy and physics.
Approximately every 10 days, we scan the literature added to the ADS since 
the last update and create custom lists of recent papers for each
subscriber, formatted to allow quick reading and access.  Subscribers
are notified by e-mail in html format.  
One can have separate notifications for the
different ADS databases and daily and/or weekly
notification for the arXiv e-print database in collaboration with the
arXiv e-print server.

In addition, we have other features to offer
the more advanced user.  For example, we provide the capability to turn
any ADS query into an RSS feed by clicking the RSS link at the bottom of a
results list.  Users can then use an RSS reader such as myYahoo or Mozilla
Firefox to read results from that query on a regular basis.  Users may also 
find it helpful to use our private library feature to group together articles 
that they commonly use or reference.  Private libraries are available at unique
URLs so that they can be shared with colleagues.

We also find that a number of librarians regularly search the ADS with very
complicated queries to try to isolate papers about topics or telescopes
particular to their institutions.  Feedback with the ADS staff may help to
fine-tune these queries, and it is also good practice to
disable synonyms for individual words such as acronyms, which may have
alternative meanings as stand-alone words.

\subsection{Searching Tips}
\small
\begin{itemize}
\item Use Full Name:  Since January 2006, the default has been to use full
author first and middle names as opposed to truncating at the author's
first initial.  
\item First Author Only:  Use a caret to get only articles where an author is
the first author of the paper, ``\^{}last name[, first name]".
\item Publication Month:  Omit month whenever possible so that unknown months
(listed as ``00") are not excluded.
\item Object Searching:  Include in SIMBAD object box {\it and} in abstract
text field (which searches text and title) to maximize results.
\item Journal Selection:  Use the Filters Section of the Main Query Form
to select or deselect specific publications, as well as to limit to refereed
publications.
\item Disabling a Synonym:  To disable a synonym for a single word, 
prepend an equal sign ``=" to the word you wish
to match exactly, (e.g. =reddening, if you want to exclude abstracts using the
word red).
\end{itemize}
\normalsize
\subsection{Additional Searching Possibilities}

Two other search capabilities deserve mentioning as they are not commonly used,
but have the potential to be very important for those tracking papers by authors
at a given institution.  First, on the main query page we offer the capability 
of selecting bibliographic records which are within a specified ``group," 
where the group may be defined as papers by 
researchers at a given institution or papers using data from a given 
telescope.  We primarily enlist the help of institutional librarians in
maintaining these groups, which enable scientists to make institute-wide 
searches easily, as well as to make bibliometric compilations trivial.

In addition, we offer a basic affiliation search which we
have not integrated into the main query form because 
affiliations found in the ADS databases are inconsistently formatted, contain 
a lot of noise, and most importantly only exist for about half of the entries 
in the database.  This means that a search by affiliation generates very 
biased results.  However, given the number of requests we have had on this 
subject, we have created a separate query form allowing a user to 
search for different affiliation spellings in the database and subsequently 
retrieve any records containing them.  That form is available at
http://adsabs.harvard.edu/list\_aff.html.   
Note however, that because of the 
limitations of this type of search, we continue to recommend that people use 
author searches when compiling bibliometric studies for particular 
institutions until we are able to find the manpower or collaborators needed
to improve this service.

\section{Classification Issues in the ADS}

As journal articles are incorporated into the ADS, there are sometimes
decisions to be made as to how these articles should be classified.
Articles are classified into separate databases to allow for 
discipline-specific searching, and articles are classified as refereed or 
non-refereed so that users have the ability to discern between the two.

\subsection{Classification of Articles into Separate Databases}

The classification of articles into separate databases in the ADS is
currently done on a journal-by-journal basis for most journals.
For journals which span multiple disciplines,
such as Science, Nature, and Publications of the National Academy of
Sciences (PNAS), we use keywords provided by the journal to decide
where to index the articles.  However, this does not correctly classify
all articles, as keywords are not always correctly anticipated, and
some journals are not able to provide us with accurate keywording.
Furthermore, this method does not work for some journals which publish
across disciplines, such as physics journals which occasionally publish 
special astronomy conferences.

We found that we needed a solution which allowed us to automate classification
so that material currently in one database can additionally be included in a more 
appropriate database.  This would give us the ability to find
material already indexed in one database which should be included
in a different database.

As a result, we have created a classification tool which uses the Abstract
Service to generate a score of how that abstract ranks against each database.
Once parameters are adjusted (such as the minimum number of words, the
weighting of certain words, and the weighting of citations from core
journals), the classifier computes a score indicating how relevant the input 
article is to each of the ADS databases and assigns the article to the database 
with the highest score.

We expect to use this tool to check the relevance of all articles in the
Physics and General Science databases.   Fine-tuning is still
in progress so that individual titles do not need to be monitored before
we can run it over large numbers, but we expect to be able to use it to
improve the division of databases by content.

\subsection{Classification of Articles into Refereed vs. Non-Refereed}

The classification of articles into refereed versus non-refereed status
in the ADS is currently done largely by hand.  Based on our knowledge
of the status of a given journal, together with input gathered both from
journal editors and librarians, we attempt to declare a status of
either refereed or non-refereed.  There are several problems with this 
approach, the biggest ones being: (1) it is subjective -- journal editors tend 
to believe their journals are refereed while librarians or scientists may 
not agree.  We should not be making the final decision; (2) it is 
time-consuming -- doing the work by hand involves researching any 
questionable conferences by hand, sending emails, asking editors for 
clarification on refereeing status; and (3) it may blur the importance 
of the qualifier ``refereed" -- refereed journals are publishing 
conference proceedings either in their main journal or as a supplement. 
Are these refereed to the same standard?

Users and librarians would like us to be strict in our definition 
of what qualifies as a refereed paper, but many editors believe that any 
refereeing process at all qualifies a paper as being refereed.  Is there
more than one level of refereeing standard?  If so, how could the ADS apply
this?  Nature, for example, is examining an alternative model of
an open peer-review process for their articles.

\section{Conclusion}

Because so many people use the ADS on such a regular basis, it is a difficult 
task to convince people to spend time learning how to improve their use of the
ADS.  When we attend conferences, we find that most people do not spend the
time to stop by our booth, telling us as they walk by that they 
``use us all the time."  Therefore, we find it difficult to spread word to 
the community that small changes may greatly improve their search results. 
Since the default searching works well for the majority
of users, we have tried instead to concentrate our efforts on improving the 
default searching, improving the data, and creating services that will 
generate results that the users desire, with minimal effort required by 
the user.  Based on feedback from users, we believe the majority of them are
satisfied, therefore we will continue to channel our efforts in these 
directions.

%\section*{}    %%% Unnumbered top level section head (remove "%" symbol)
%\subsection*{}   %%% Unnumbered second level section head (remove "%" symbol)

% \acknowledgements %%% Text of acknowledgements runs on after this command.

%%% THE BIBLIOGRAPHY
%%%
%%% CONSULT SECTION 3 OF "INSTRUCTIONS FOR AUTHORS" FOR HOW TO USE NATBIB.
%%% AUTHORS ARE ENCOURAGED TO USE EITHER THE "THEBIBLIOGRAPY" ENVIRONMENT
%%% BY UNCOMMENTING (DELETING THE "%" SYMBOL) THE COMMANDS BELOW, OR BY
%%% USING THE BIBTEX ENVIRONMENT. TO FIND OUT WHICH IS APPLICABLE TO YOUR
%%% CONTRIBUTION, CONSULT THE VOLUME EDITORS FOR YOUR PROCEEDINGS.
%%%

%\begin{thebibliography}{}
%\bibitem[]{}
%\bibitem[]{}
%\bibitem[]{}
%\bibitem[]{}
%\bibitem[]{}
%\bibitem[]{}
%\bibitem[]{}
%\bibitem[]{}
%\bibitem[]{}
%\bibitem[]{}
%\bibitem[]{}
%\bibitem[]{}
%\end{thebibliography}

\end{document}